\documentclass[graybox]{svmult}

\usepackage{mathptmx}       % selects Times Roman as basic font
\usepackage{helvet}         % selects Helvetica as sans-serif font
\usepackage{courier}        % selects Courier as typewriter font
\usepackage{type1cm}        % activate if the above 3 fonts are
\usepackage{makeidx}         % allows index generation
\usepackage{graphicx}        % standard LaTeX graphics tool
\usepackage{multicol}        % used for the two-column index
\usepackage[bottom]{footmisc}% places footnotes at page bottom

\usepackage{amsmath}
\usepackage{amsfonts}
\usepackage{amssymb}
\usepackage{graphics}
\usepackage{epsfig}
\usepackage{subfigure}

\usepackage{rotating}
\usepackage{multirow}
\usepackage{pdflscape}
\usepackage{scalefnt}

\makeindex             % used for the subject index
\begin{document}

\title*{Multilayer networks: metrics and spectral properties}
\author{Emanuele Cozzo, Guilherme Ferraz de Arruda, Francisco A. Rodrigues and Yamir Moreno}
\institute{Emanuele Cozzo \at Institute for Biocomputation and Physics of Complex Systems (BIFI) and Department of Theoretical Physics, University of Zaragoza, 50018 Zaragoza, Spain, \email{emcozzo@gmail.com}
\and Guilherme Ferraz de Arruda \at Departamento de Matem\'{a}tica Aplicada e Estat\'{i}stica, Instituto de Ci\^{e}ncias Matem\'{a}ticas e de Computa\c{c}\~{a}o,
Universidade de S\~{a}o Paulo - Campus de S\~{a}o Carlos, Caixa Postal 668,
13560-970 S\~{a}o Carlos, SP, Brazil. \email{gui.f.arruda@gmail.com}
\and  Francisco A. Rodrigues \at Departamento de Matem\'{a}tica Aplicada e Estat\'{i}stica, Instituto de Ci\^{e}ncias Matem\'{a}ticas e de Computa\c{c}\~{a}o,
Universidade de S\~{a}o Paulo - Campus de S\~{a}o Carlos, Caixa Postal 668,
13560-970 S\~{a}o Carlos, SP, Brazil. \email{francisco.rodrigues.usp@gmail.com}
\and Yamir Moreno \at Institute for Biocomputation and Physics of Complex Systems (BIFI) and Department of Theoretical Physics, University of Zaragoza, 50018 Zaragoza, Spain and Complex Networks and Systems Lagrange Lab, Institute for Scientific Interchange, Turin, Italy \email{yamir.moreno@gmail.com}}
\maketitle

\abstract{Multilayer networks represent systems in which there are several topological levels each one representing one kind of interaction or interdependency between the systems' elements. These networks have attracted a lot of attention recently because their study allows considering different dynamical modes concurrently. Here, we revise the main concepts and tools developed up to date. Specifically, we focus on several metrics for multilayer network characterization as well as on the spectral properties of the system, which ultimately enable for the dynamical characterization of several critical phenomena. The theoretical framework is also applied for description of real-world multilayer systems.}

\section{Introduction}

Complex network science relies on the hypothesis that the behavior of many complex systems can be explained by studying structural and functional relations among its components by means of a graph representation. The emergence of interconnected network models responds to the fact that complex systems include multiple subsystems organized as layers of connectivity. In this way, interconnected networks have emerged during the last few years as a general framework to deal with hyperconnected systems~\cite{vespignani}. With the term interconnected networks one may refer to many types of connections among different networked systems: dependency relations among systems of different objects, cooperative or competitive relations among systems of different agents, or different channels of interactions among the same set of actors, to name a few. What these examples have in common is that different interaction modes among a differentiated or indistinguishable set of components/actors might exist.

Although this framework has been used for many years, only in the last several years it has attracted more attention and a number of formalisms have been proposed to deal with multilayer networks~\cite{tensorpaper,boccaletti}. Here we elaborate on a formalism developed recently and discussed at length in the review paper by Kivela et al. \cite{mikko}. To this end, we report on a more refined formalism that is aimed at optimizing the study of a particular case of interconnected networks that is of much interest: Multiplex Networks.

In Multiplex Networks a set of agents might interact in different ways, i.e., through different means. Since a subset of agents is present at the same time in different networks of interactions (layers), these layers become interconnected. Examples of such type of systems can be founded in different fields, from biological systems, where the web of molecular interactions in a cell make use of many different biochemical channels and pathways, to technological systems, where person-to-person communication (usually machine-mediated) happens across many different modes. We take the last example as a paradigmatic one, which gave rise to the now popular term "hyperconnectivity"~\cite{wellman}.

Suppose  we are interested in analyzing a set of social agents (individuals, institutions, firms, etc. ), who interact among them through a number of online social networks (OSNs) like Twitter, Facebook, etc. Some of these agents might be present in several OSNs and exchange information through them, using the information obtained in one network to communicate in another one, or integrating information across all of those in which they are active. We represent such a system as a set of graphs, one for each OSN, in which each actor who participates in it is represented by a node. These networks are the layers of the graph. In this scheme, the same actors are represented by a number of different nodes (as many nodes as the number of layers in which the actor is present). At the same time, we represent the fact that different nodes might denote the same actors, thus being related, by a coupling graph in which nodes representing the same actors are connected. 

The rest of the Chapter is organized as follows. The first section translates the aforementioned structural features in the formal language of graph theory. By doing that, we synthesize the topology of such a system in terms of matrices. In addition, as many years of research \cite{boccalettireview} have demonstrated, the relation between structure and function can be studied by means of the spectral properties of the matrices representing the graph structure. This is also studied in the second part of this Chapter, where we give a simple example of the epidemic spreading process and analyze real world multilayer networks.

\section{Notation, basic definitions and properties}

A multiplex network is a quadruple $\mathcal{M}=(\mathfrak{L},\mathfrak{n},\mathfrak{P},\mathfrak{M})$. $\mathfrak{L}=\{1,\dots,m\}$ is an index set that we call the layer set. Here we have assumed $\mathfrak{L}\subset\mathcal{N}$ for practical reasons and without loss of generality. We indicate the general element of $\mathfrak{L}$ with Greek lower case letters. Moreover, $\mathfrak{n}$ is a set of nodes and $\mathfrak{P}=(\mathfrak{n},\mathfrak{L},\mathfrak{N})$,  $\mathfrak{N}\subseteq\mathfrak{n}\times\mathfrak{L}$ is a binary relation. Finally, the statement $(n,\alpha)\in \mathfrak{N}$ is read \textit{node n participates in layer $\alpha$}. We call the ordered pair $(n,\alpha)\in\mathfrak{N}$ a node-layer pair and we say that the node-layer pair $(n,\alpha)$ is the representative of node $n$ in layer $\alpha$.

On the other hand, $\mathfrak{M}=\{G_\alpha\}_{\alpha\in\mathfrak{L}}$ is a set of graphs, that we call layer-graphs, indexed by means of $\mathfrak{L}$. The node set of a layer-graph $G_\beta\in \mathfrak{M}$ is a sub-set $\mathfrak{n}_\beta\subset\mathfrak{N}$ such that $\mathfrak{n}_\beta=\{(n,\alpha)\in\mathfrak{P}\mid \alpha=\beta\}$, so the nodes of $G_\beta$ are node-layer pairs; in that sense we say that node-layer pairs represent nodes in layers. The edge set of a graph $G_\alpha\in\mathfrak{M}$ is $\mathfrak{E}_\beta\subseteq \mathfrak{n}_\beta\times \mathfrak{n}_\beta$. Additionally, the binary relation $\mathfrak{P}$ can be identified with its graph $G_\mathfrak{P}$. $G_\mathfrak{P}$ has nodes set given by $\mathfrak{n}\cup \mathfrak{L}$, and edge set $\mathfrak{E}_\mathfrak{P}=\mathfrak{N}$,  and we call it the \textit{participation graph}.

Consider the graph $G_\mathfrak{C}$ on $\mathfrak{N}$ in which there is an edge between two node-layer pairs $(n,\alpha)$ and $(m,\beta)$ only if $n=m$; that is, only if the two edges in the graph $G_\mathfrak{P}$ are incident on the same node $n\in\mathfrak{n}$, which means that the two node-layer pairs represent the same node in different layers. We call $G_\mathfrak{C}$ the coupling graph. It is easy to realize that the coupling graph is formed by $n=\mid\mathfrak{n}\mid$ disconnected components that are clicks or isolated nodes. Each clique is formed by all the representatives of a node in the layers, we call the components of $G_\mathfrak{C}$ \textit{supra-nodes}.

Let's now also consider the graph $G_\mathfrak{l}$ on the same nodes set $\mathfrak{N}$, and in which there is an edge between two node-layer pairs $(n,\alpha)$, $(m,\beta)$ only if $\alpha=\beta$; that is, only if the two edges in the graph $G_\mathfrak{P}$ are incident on the same node $\alpha\in\mathfrak{L}$. We call $G_\mathfrak{l}$ the layer graph. It is easy to realize that graph is formed by $m=\mid\mathfrak{L}\mid$ disconnected components that are clicks.

Finally, we can define the \textit{supra-graph} $G_\mathcal{M}$ as the union of the layer-graphs with the coupling graph: $G_\mathcal{C}\cup \mathfrak{M}$. $G_\mathcal{M}$ has node set $\mathfrak{N}$ and edge set $\bigcup_\alpha \mathfrak{E}_\alpha \cup \mathfrak{E}_\mathfrak{C}$. $G_\mathcal{M}$ is a synthetic representation of the Multiplex Network $\mathcal{M}$. It results that each layer-graph $G_\alpha$ is a sub-graph of $G_\mathcal{M}$ induced by $\mathfrak{n}_\alpha$. Furthermore, when all nodes participate in all layer-graphs the Multiplex Network is said to be fully aligned \cite{mikko} and the coupling graph is made of $n$ complete graphs of $m$ nodes.

It is useful to come back to our system of social agents as a paradigmatic multiplex network to make sense of the previous definitions. The layer set is the list of OSNs, for example $\mathfrak{L}=\{Facebook,Twitter,Google+\}$. Since for practical purposes we want a set of indexes that are natural numbers, we may say that: Facebook is $1$, Twitter is $2$, and $Google+$ is 3. The set of nodes is the set of social actors, for example $n=\{Marc, Alice, BiFi, Nick, Rose\}$. The binary relations represent the participation of each of these agents in some of the OSNs, thus we have that an statement of the type \textit{Alice has a Facebook account} is represented by the pair $(Alice,1)$, that is a node-layer pair. Each set of relation in each OSN is represented by a graph, for example the link $[(Alice,1),(Nick,1)]$ means that Alice and Nick are friends on Facebook. If Alice has a Facebook account and a Twitter account, but not a Google+ account, in the coupling graph we will have the connected component $[(Alice,1),(Alice,2)]$ that is the supra-node related to Alice. If only the BiFi, Nick, and Rose have Google+ accounts, in the layer graph we will have the connected component $[(Bifi,3),(Nick,3),(Rose,3)]$.

\section{Multiplex networks related Matrices}
\subsection{Adjacency matrices}

In general, the adjacency matrix of a (unweighted, undirected) graph $G$ with $N$ nodes is a $N\times N$ (symmetric) matrix $\mathbf{A}=\{a_{ij}\}$, with $a_{ij}=1$ only if there is an edge between $i$ and $j$ in $G$, and $a_{ij}=0$ otherwise. We can consider the adjacency matrix of each of the graphs introduced in the previous section. The adjacency matrix of a layer graph $G_\alpha$ is a $n_\alpha\times n_\alpha$ symmetric matrix $\mathbf{A}^\alpha={a^\alpha_{ij}}$, with $a^\alpha_{ij}=1$ only if there is an edge between $(i,\alpha)$ and $(j,\alpha)$ in $G^\alpha$. We call them layer adjacency matrices.

\begin{figure}[H]
\begin{center}
\includegraphics[width=10cm]{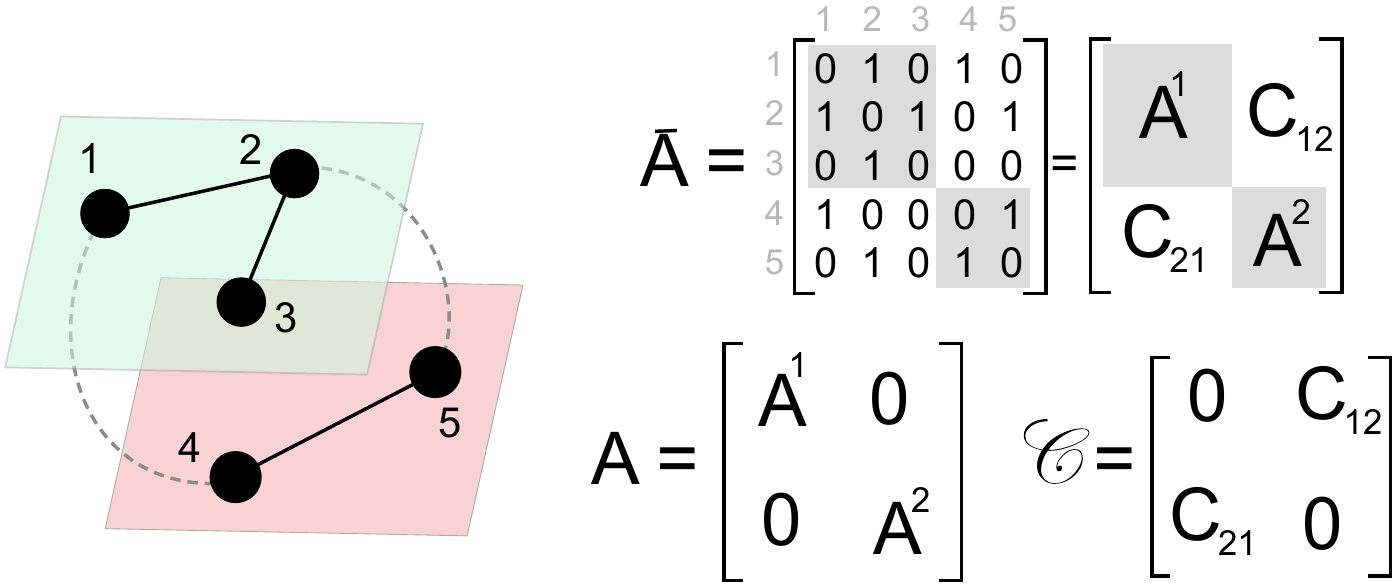}
\end{center}
\caption{Example of a multiplex network. The structure of each layer is represented by an adjacency matrix  $\mathcal{A}^i$, where $i=$1, 2. $\mathcal{C}_{lm}$ stores the connections between layers $l$ and $m$. Note that the number of nodes in each layer is not the same. }
\label{fig:Matrices}
\end{figure}

Likewise, the adjacency matrix of $G_\mathfrak{P}$ is an $n\times m$ matrix $\mathcal{P}=p_{i\alpha}$, with $p_{i\alpha}=1$ only if there is an edge between the node $i$ and the layer $\alpha$ in the participation graph, i.e. only if node $i$ participate in layer $\alpha$. We call it the participation matrix. The adjacency matrix of the coupling graph $G_\mathfrak{C}$ is an $N \times N$ matrix $\mathcal{C}=\{c_{ij}\}$, with $c_{ij}=1$ only if there is an edge between node-layer pair $i$ and $j$  in $G_\mathfrak{C}$, i.e. if they are representatives of the same node in different layers. We can arrange the rows and the columns of $\mathcal{C}$ such that node-layer pairs of the same layer are contiguous and layers are ordered. We assume that $\mathcal{C}$ is always arranged in that way. It results that $\mathcal{C}$ is a block matrix with zero diagonal blocks. Thus, $c_{ij}=1$, with $i,j=1,\dots , N$ represents an edge between a node-layer pair in layer $1$ and a node-layer pair in layer $2$ if $i<n_1$ and $n_1<j<n_2$. Figure~\ref{fig:Matrices} shows a multiplex network and the respective matrices $A$ and $\mathcal{C}$.

\subsection{The supra-adjacency matrix}

The  \textit{supra-adjacency} matrix is the adjacency matrix of the supra-graph $G_\mathcal{M}$. Just as $G_\mathcal{M}$, $\bar{\mathcal{A}}$ is a synthetic representation of the whole multiplex $\mathcal{M}$. By definition, it can be obtained from the intra-layer adjacency matrices and the coupling matrix in the following way:
\begin{equation}
\bar{\mathcal{A}}=\bigoplus_\alpha \mathbf{A}^\alpha + \mathcal{C},
\label{supradjacency}
\end{equation}  
where the same consideration as in $\mathcal{C}$ applies for the indices. We also define $\mathcal{A}=\bigoplus \mathbf{A}^\alpha$, and we call it the intra-layer adjacency matrix. Figure~\ref{fig:Matrices} shows the supra-adjacency and the intra-layer adjacency matrices of a multiplex network. Some basic metrics are easily calculated from the supra-adjacency matrix. 

The degree of a node-layer $i$ is the number of node-layers connected to it by an edge in $G_\mathcal{M}$ and is given by
\begin{equation}
K_{i}=\sum_j \bar{\mathcal{A}}_{ij}.
\label{degree}
\end{equation}
Sometimes we write $i(\alpha)$ as an index, instead of simply $i$, to explicitly indicate that the node-layer $i$ is in layer $\alpha$ even if the index $i$ already uniquely indicates a node-layer pair. Since $\bar{\mathcal{A}}$ can be read as a block matrix, with the $\mathbf{A}^\alpha$ on the diagonal blocks, the index $i(\alpha)$ can be interpreted as block index. It is also useful to define the following quantities
\begin{equation}
e_\alpha=\sum_{\beta<\alpha}n_\beta,
\label{excess}
\end{equation}
which we call the excess index of layer $\alpha$. The layer degree of a node-layer $i$, $k_{i(\alpha)}$, is the number of neighbors it has in $G^\alpha$, i.e., $k_{i(\alpha)}=\sum_j a^\alpha_{ij}$. By definition of $\bar{\mathcal{A}}$
\begin{equation}
k_{i(\alpha)}=\sum_{j=1+e_\alpha}^{n_\alpha +e_\alpha}\bar{\mathcal{A}}
_{ij}.
\label{intradegree}
\end{equation}
The coupling degree of a node-layer $i$, $c_{i(\alpha)}$, is the number of neighbors it has in the coupling graph, i.e., $c_{i(\alpha)}=\sum_j c_{ij}$. From $\bar{\mathcal{A}}$ we get

\begin{equation}
c_{i\alpha}=\sum_{\substack{j<e_\alpha,\\ j>n_\alpha +e_\alpha}}\bar{\mathcal{A}}_{ij}.
\label{couplindegree}
\end{equation}
Finally, we note that the degree of a node-layer can be expressed as
\begin{equation}
K_{i(\alpha)}=\sum_j \bar{\mathcal{A}}_{ij}=k_{i\alpha}+c_{i\alpha}.
\label{degree2}
\end{equation}
Eq.(\ref{degree2}) explicitly expresses the fact that the degree of a node-layer pair is the sum of its layer-degree plus its coupling-degree. 

\subsection{The supra-Laplacian matrix}
Generally, the Laplacian matrix of a graph with adjacency matrix $\mathbf{A}$, or simply the Laplacian, is given by
\begin{equation}
\mathcal{L}=\mathcal{D}-\mathcal{A}
\label{laplacian}
\end{equation}
where $\mathbf{D}=diag(k_1,k_2,\dots)$ is the degree matrix.

Thus, it is natural to define the \textit{supra-Laplacian} matrix of a Multiplex network as the Laplacian of its supra-graph
\begin{equation}
\bar{\mathcal{L}}=\bar{\mathcal{D}}-\bar{\mathcal{A}},
\label{supralaplacian}
\end{equation} 
where $\bar{\mathcal{D}}=diag(K_1,K_2,\dots,K_N)$ is the degree matrix. Besides, we can define the layer Laplacian of each graph $G_\alpha$ as
\begin{equation}
\mathbf{L}_\alpha=\mathbf{D}^\alpha -\mathbf{A}^\alpha,
\end{equation} 
and the Laplacian of the coupling graph
\begin{equation}
\mathcal{L}_C=\Delta-\mathcal{C}
\end{equation}
where $\Delta=diag(c_1,c_2,\dots,c_N)$ is the coupling-degree matrix.\\
By definition, we have 
\begin{equation}
\bar{\mathcal{L}}=\bigoplus_\alpha \mathcal{L}^\alpha +\mathcal{L}_C.
\label{suprasum}
\end{equation}
Eq. (\ref{suprasum}) takes a very simple form in the case of a node-aligned multiplex, i.e.,
\begin{equation}
\bar{\mathcal{L}}=\bigoplus_\alpha (\mathbf{L}^\alpha +c I_N) -\mathbf{K}_m \otimes I_n 
\end{equation}
where $\mathcal{K}_m$ is the adjacency matrix of a complete graph of $m$ nodes, $I_k$ is the $k\times k$ identity matrix and $c_i=c, \forall i\in \mathfrak{N}$ is the coupling degree of a node-layer pair.

\subsection{Characteristic Matrices}
\subsubsection{Supra-nodes Characteristic Matrix}

The supra-nodes characteristic matrix $\mathcal{S}_\mathfrak{n}=\{s_{ij}\}$ is an $N\times n$ matrix with $s_{ij}=1$ only if the node-layer $i$ is a representative of node $j$, i.e., it is in the connected component $j$ in the graph $G_\mathfrak{C}$. We call it a characteristic matrix since supra-nodes partitions the node-layer set and $S_n$ is the characteristic matrix of that partition.

\subsubsection{Layers Characteristic Matrix}
The layer characteristic matrix $\mathcal{S}_\mathfrak{l}=\{s_{ij}\}$ is an $N\times m$ matrix with $s_{ij}=1$ only if the node-layer $i$ is in the connected component $j$ in the graph $G_\mathfrak{l}$. We call it a characteristic matrix since it is the characteristic matrix of the partition of the node-layer set induced by layers.

\section{The coarse-grained representation of a Multiplex Network}
\subsection{Nodes partitions and Quotient graphs}
We next briefly introduce the notion of network quotient associated to a partition of the node set. Suppose that ${V_1, . . . ,V_m}$ is a partition of the node set of a
network $G$ with adjacency matrix $A$, and write $n_i = |V_i|$. The quotient network $Q$ of $G$ is a coarse-grained representation of the network with respect to the partition. It has one node per cluster $V_i$ and
an edge from $V_i$ to $V_j$ weighted by an average connectivity
from $V_i$ to $V_j$
\begin{equation}
b_{ij}=\frac{1}{\sigma}\sum_{\substack{k\in V_i\\l\in V_j}}a_{kl}.
\label{quotientelement}
\end{equation}
Different choices are possible for the normalization parameter $\sigma$: $\sigma_i=n_i,\ \sigma_j=n_j$ or $\sigma_{ij}=\sqrt{n_in_j}$. Depending on the choice for $\sigma$ we call the resulting quotient respectively: left, right or symmetric quotient. We can express the left quotient $Q_l(A)$ in matrix form. Consider the $n\times m$ characteristic matrix of the partition $S=s_{ij}$, with $s_{ij}=1$ if $i\in V_j$ and zero otherwise. Then
\begin{equation}
Q_l(A)=\Lambda^{-1}S^TAS,
\label{quotientadjacency}
\end{equation} 
where $\Lambda=diag\{n_1,\dots,n_m\}$.

\subsection{Aggregate Network and Network of Layers of a Multiplex Network}
In the context of Multiplex Networks two quotient graphs arise naturally \cite{ruben} by considering coupled node-layer pairs and layers. Supra-nodes partition the supra-graph, and the supra-nodes characteristic matrix $S_n$  is the associated characteristic matrix. Then, we define the aggregate network of the multiplex network as the quotient associated to that partition:
\begin{equation}
\tilde{\mathbf{A}}=\Lambda^{-1}\mathcal{S}_n^T\bar{\mathcal{A}}\mathcal{S}_n,
\label{aggregate}
\end{equation}  
where $\Lambda=diag\{\kappa_1,\dots,\kappa_n\}$ is the multiplexity degree matrix.
Since, the Laplacian of the quotient is equal to the quotient of the Laplacian, the Laplacian of the aggregate network is given by:
\begin{equation}
\tilde{\mathbf{L}}=\Lambda^{-1}\mathcal{S}_n^T\bar{\mathcal{L}}\mathcal{S}_n.
\label{aggregateLaplacian}
\end{equation}
In the same way, layers partition the supra-graph, thus the network of layers is defined by
\begin{equation} \label{network_of_layers}
\tilde{\mathbf{A}}_\mathfrak{l}=\Lambda^{-1}\mathcal{S}_\mathfrak{l}^T\bar{\mathcal{A}}\mathcal{S}_\mathfrak{l},
\end{equation}
and its Laplacian is given by
\begin{equation}
\tilde{L}_\mathfrak{l}=\Lambda^{-1}S_\mathfrak{l}^T\bar{L}S_\mathfrak{l}.
\label{nolLaplacian}
\end{equation}

\section{Spectral Properties}
\subsection{The largest eigenvalue of $\bar{\mathcal{A}}$}
\begin{figure}[H]
\begin{center}
\includegraphics[width=10cm]{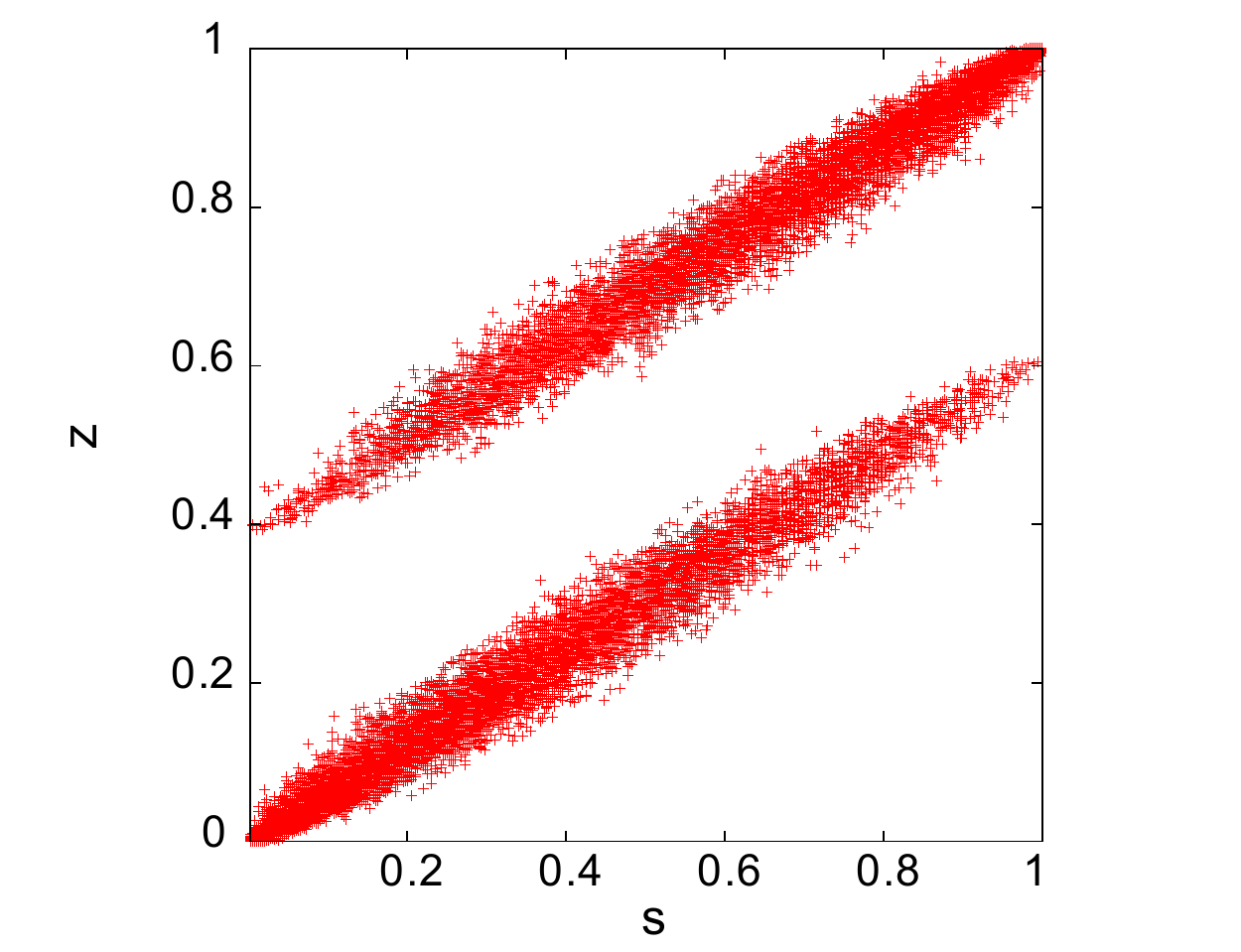}
\includegraphics[width=10cm]{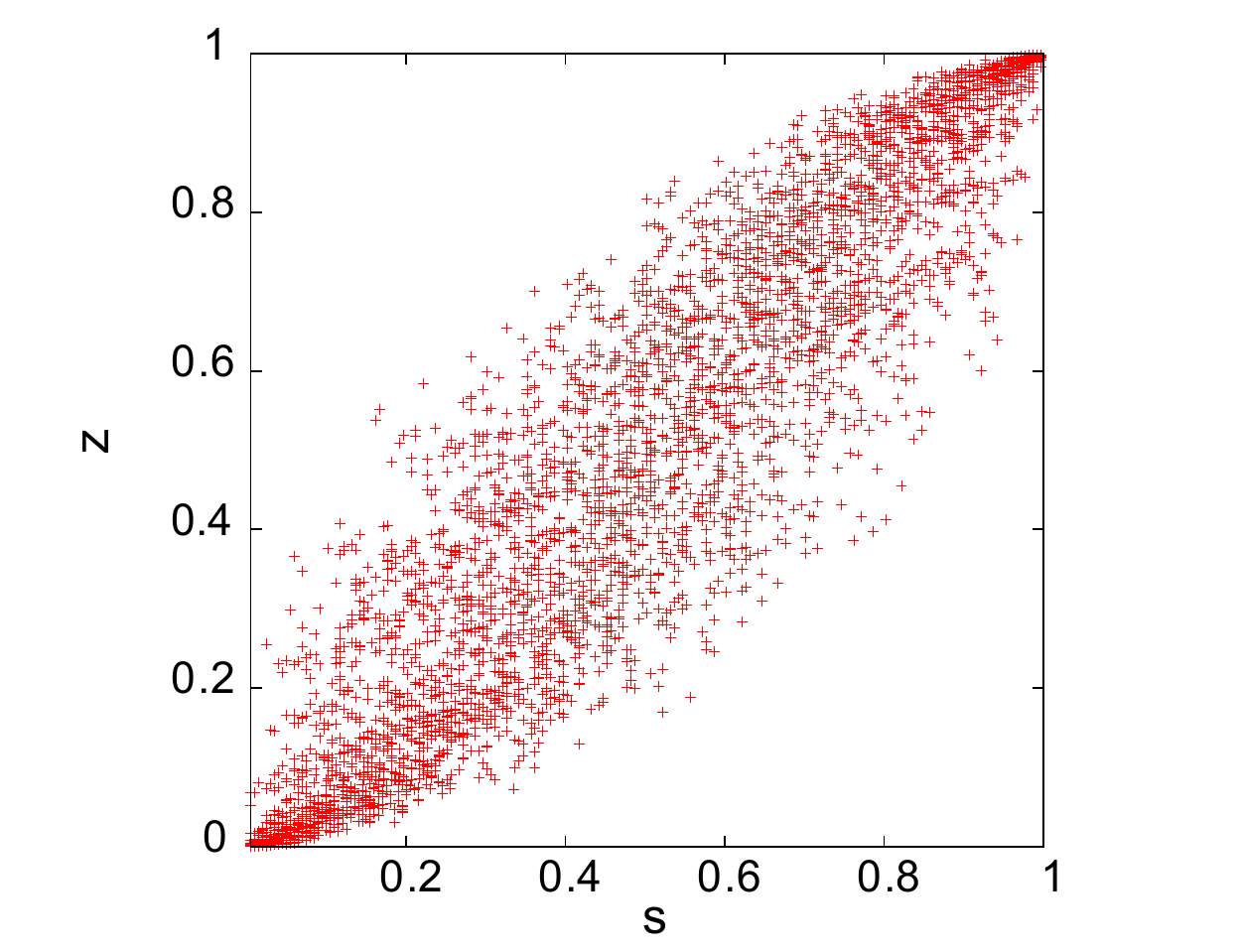}
\end{center}
\caption{Effective multiplexity $z$ as a function of the fraction of nodes coupled $s$ for a two layers multiplex with 800 nodes with a power law distribution with $\gamma=2.3$ in each layer. For each value of $s$, 40 different realizations of the coupling are shown while the intra-layer structure is fixed. In the panel on the top the $z$ shows a two band structure, while in the panel on the bottom, it is continuous. The difference is due to the structure of the eigenvector.}
\label{fig1}
\end{figure}

\begin{figure}[H]
\begin{center}
\includegraphics[width=10cm]{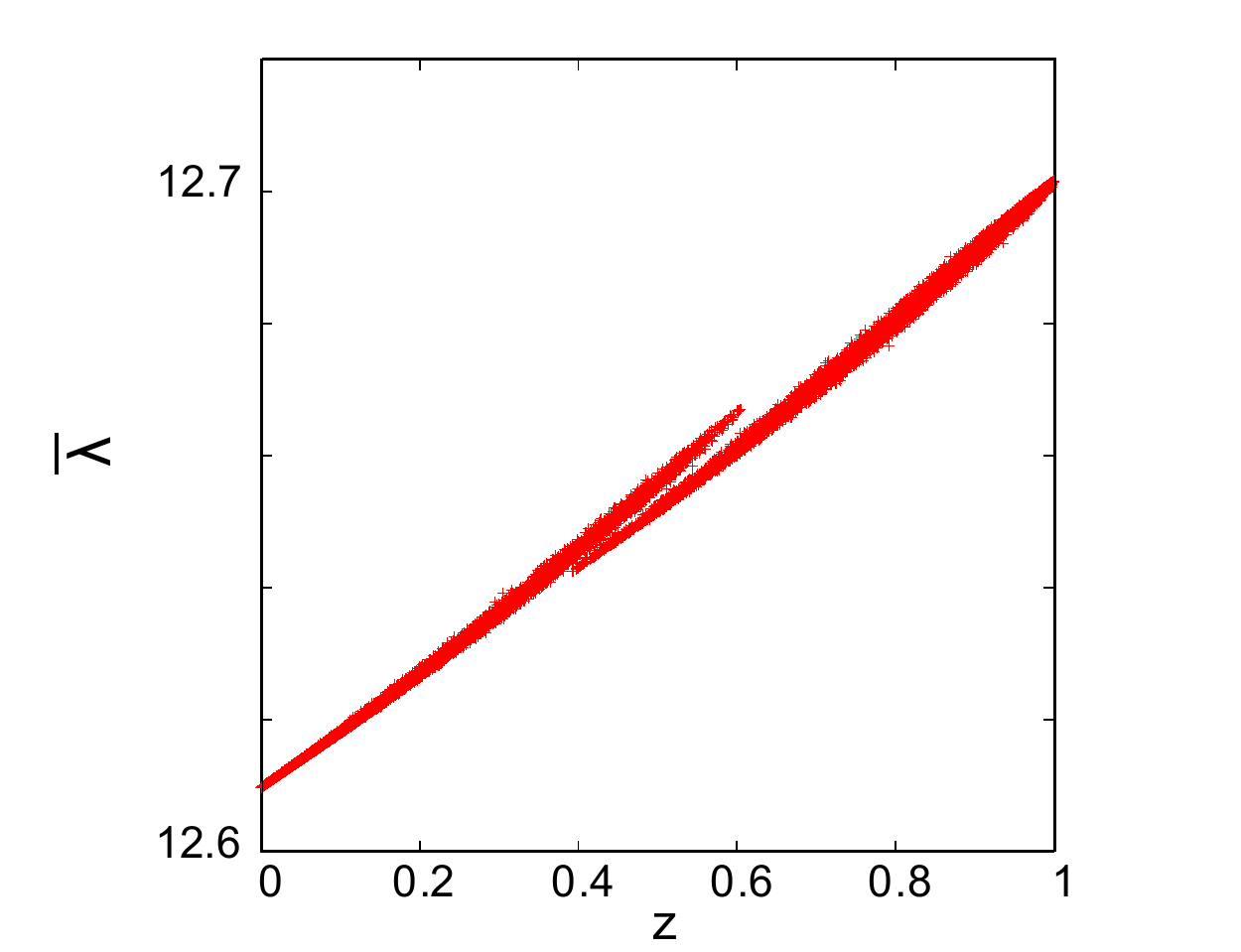}
\includegraphics[width=10cm]{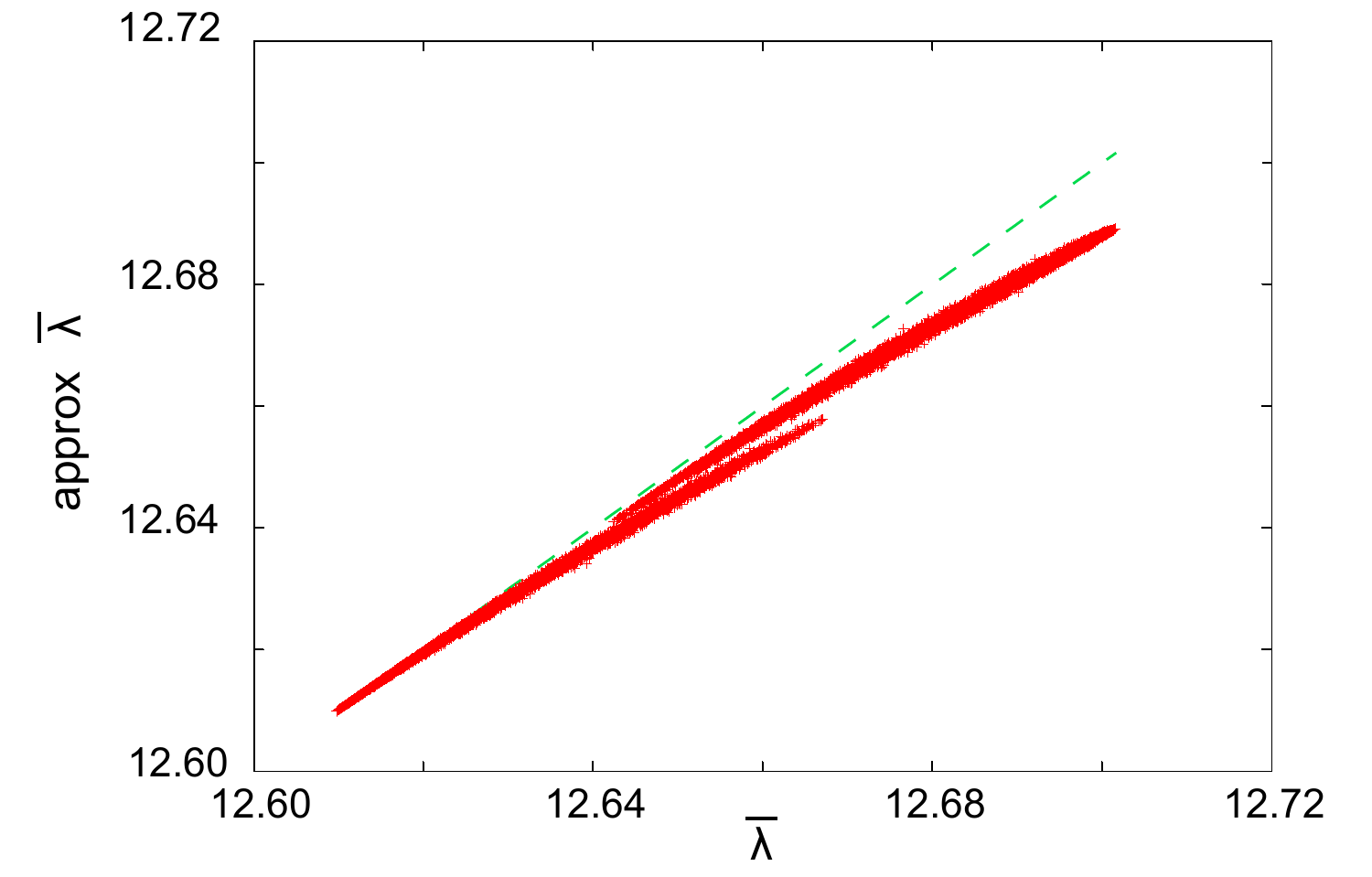}
\end{center}
\caption{Same setting of top panel of previous figure. On the top: calculated $\bar{\lambda}$. We can see two branches corresponding to the two branches of the previous figure. Bottom: calculated vs approximated $\bar{\lambda}$}
\label{fig2}
\end{figure}

In the following we will interpret $\bar{\mathcal{A}}$ as a perturbed version of $\mathcal{A}$, $\mathcal{C}$ being the perturbation. This choice is reasonable whenever
\begin{equation}
\mid\mid \mathcal{C}\mid\mid<\mid\mid \mathcal{A}\mid\mid.
\end{equation}
Consider the largest eigenvalue $\lambda$ of $\mathcal{A}$. Since $\mathcal{A}$ is a block diagonal matrix, the spectrum of $\mathcal{A}$, $\sigma(\mathcal{A})$, is
\begin{equation}
\sigma(\mathcal{A})=\bigcup_{\alpha}\sigma(\mathbf{A}^\alpha),
\end{equation} 
$\sigma(\mathbf{A}^\alpha)$ being the spectrum of the adjacency-matrix $\mathbf{A}^\alpha$ of layer $\alpha$. So, the largest eigenvalue $\lambda$ of $\mathcal{A}$ is
\begin{equation}
\lambda = \max_\alpha \lambda_\alpha
\end{equation}
with $\lambda_\alpha$ being the largest eigenvalue of $\mathbf{A}^\alpha$. We will look for the largest eigenvalue $\bar{\lambda}$ of $\bar{\mathcal{A}}$ as
\begin{equation}
\bar{\lambda}=\lambda + \Delta\lambda,
\end{equation}
where $\Delta\lambda$ is the perturbation to $\lambda$ due to the coupling $C$. For this reason, we call the layer $\delta$ for which $\lambda_\delta=\lambda$ the dominant layer. Let $\mathbf{1}_\alpha$ be a vector of size $m$ with all entries equal to $0$ except for the $\delta$-th. If $\phi_\delta$ is the eigenvector of $\mathbf{A}^\delta$ associated to $\lambda_\delta$, we have that 
\begin{equation} \label{phi}
\phi=\phi_\delta\otimes\mathbf{1}_\alpha
\end{equation}
is the eigenvector associated to $\lambda$. Observe that $\phi$ have dimension $n_{\delta}$, while $\mathbf{1}_\alpha$ have dimension $m$, where $n_{\delta}$ is the number of nodes on the dominant layer $\delta$, yielding to a product of dimension $n_{\delta} \times m$, however it is not true if the number of nodes in is not the same on all layers. In such case we must construct the vector $\phi$ with zeros on all positions, except on the position of the leading eigenvector of the dominant layer. Then, we can approximate $\Delta\lambda$ as
\begin{equation}
\Delta\lambda\approx \frac{\phi^T \mathcal{C}\phi}{\phi^T\phi}+\frac{1}{\lambda}\frac{\phi^T \mathcal{C}^2 \phi}{\phi^T\phi}.
\label{deltaapprox}
\end{equation}
Because of the structure of $\phi$ and $C$, the first term on the \emph{r.h.s.} is zero, while only the diagonal blocks of $C^2$ take part in the product $\phi^T C^2 \phi$. The diagonal blocks of $C^2$ are diagonals and
\begin{equation}
(C^2)_{ii}=\sum_{i^{\prime}}C_{ii^\prime}C_{i^\prime i}=c_{i}.
\end{equation}
Thus, we have that the perturbation is 
\begin{equation}
\Delta\lambda\approx \frac{z}{\lambda},
\label{Iapprox}
\end{equation}
where we have defined the \emph{effective multiplexity} $z$ as the weighted mean of the coupling degree with the weight given by the squares of the entries of the leading eigenvector of $A$:
\begin{equation} \label{eq:z}
z=\sum_i c_i\frac{\phi_i^2}{\phi^T\phi},
\end{equation}
where $z = 0$ in a monoplex -single layer- network or  $z=m-1$ in a node aligned multiplex. Summing up, we have that the largest eigenvalue of the supra-adjacency matrix is equal to the largest eigenvalue of the dominant layer adjacency matrix at a first order approximation. As a consequence, for example, the critical point for an epidemic outbreak in a multiplex network is settled by that of the dominant layer at a first order approximation\cite{multispread}. At second order, the deviation of $\bar{\lambda}$ from $\lambda$ depends on the effective multiplexity and goes to zero with $\lambda$. See figure~\ref{fig1} and~\ref{fig2}.

The approximation given in Eq. (\ref{Iapprox}) can fail when the largest eigenvalue is near degenerated. We have two cases in which this can happen:
\begin{itemize}
\item the dominant layer is near degenerated,
\item there is one (or more) layers with the largest eigenvalue near that of the dominant layer. 
\end{itemize}
The accuracy of the approximation is related to the formula
\begin{equation}
\Delta \lambda \approx \phi^T \mathcal{C} \phi + \sum_{i}\frac{(\phi^{(i)T}\mathcal{C}\phi)}{\lambda-\lambda^{(i)}},
\end{equation}
where $\lambda^{(i)}$ and $\phi^{(i)}$ are the non-dominant eigenvalues and the associated eigenvectors. In the first case it is evident that the second term on the \emph{r.h.s.} will diverge, while in the latter, because of the structure of $\mathcal{C}$, $\phi$, and $\phi^{(i)}$, it is zero. In that case, we say that the multiplex network is near degenerated and we call the layers with the largest eigenvalues \emph{co-dominant layers}.

When the multiplex network is near degenerated, $\phi$ used in the approximation of equation (\ref{Iapprox}) has a different structure. Consider that we have $l$ co-dominant layers $\delta_i,\ i=1,\dots,l$. If $\phi_{\delta_i}$ is the eigenvector of $A^{\delta_i}$ associated to $\lambda_{\delta_i}$, we have that 
\begin{equation} \label{phi2}
\phi=\sum_{i=1}^l\phi_{\delta_i}\otimes\mathbf{1}_{\delta_i}.
\end{equation} 
Note that the same comment on Eq.~(\ref{phi}) also applies here. The term linear in $C$ in the approximation of equation (\ref{Iapprox}) is no more zero. We have
\begin{equation}
z_c=\frac{\phi^T \mathcal{C}\phi}{\phi^T\phi}=\frac{1}{\phi^T\phi}\sum_{l,m: l\neq m}\phi_{\delta_l}^T\phi_{\delta_m}
\end{equation}
and we name $z_c$ the \emph{correlated multiplexity}. We can decompose $z_c$ in the contribution of each single node-layer pair
\begin{equation}
{z_c}_i=\frac{1}{\phi^T\phi}\sum_{m:m\neq l}\sum_{j}{\phi_{\delta_l}}_iC_{ij}{\phi_{\delta_m}}_j.
\end{equation}
and we call ${z_c}_i$ the \emph{correlated multiplexity degree} of node-layer $i$. By definition, coupled node-layer pairs have the same correlated multiplexity degree. So, if we have $m_d$ co-dominant layers in the multiplex, we get
\begin{equation}
\Delta \lambda\approx z_c+\frac{z}{\lambda}=m_d \sum_{i\in \delta} {z_c}_i+\frac{\sum_{i\in \delta}z_i}{\lambda}.
\end{equation}

%%%%%%%%%%

%%%%%%%%%%%%
\subsection{Spectral relations between \textit{supra} and coarse-grained representations} \label{sec:spectral}
The fundamental spectral result related to a quotient network is that adjacency eigenvalues of a quotient network interlace the adjacency eigenvalues of the
parent network. That is, if $\mu_i,\dots,\mu_m$ are the adjacency eigenvalues of the quotient network, and $\lambda_i,\dots,\lambda_n$ are the adjacency eigenvalues of the parent network, it results that
\begin{equation}
\lambda_i\leq\mu_i\leq\lambda_{i+n-m}.
\label{interlacing}
\end{equation}
The same result applies for Laplacian eigenvalues. We can derive directly from that result a list of bounds for the supra-adjacency and the supra-Laplacian in terms of the aggregate network and of the network of layers \cite{ruben}. Besides, in the case of node aligned multiplex networks, we have that the eigenvalues of the laplacian of the network of layers are a sub-set of the spectrum of the supra-Laplacian. This result is of special relevance in studying the structural properties of a multiplex network, since it states that the adjacency (Laplacian) eigenvalues of the coarse-grained representation of a multiplex interlace the adjacency (Laplacian) eigenvalues of the parent. In the case of a node-aligned multiplex, the Laplacian eigenvalues of the network of layers are a sub-set of the Laplacian eigenvalues of the parent Multiplex network.

\subsection{The second eigenvalue of $\bar{L}$}
A number of structural and dynamical properties of a network can be derived from the value of the first non-zero eigenvalue of the Laplacian. In the particular case of Multiplex Networks it has been shown that its behavior reflects a structural transition of the system \cite{radicchi}. We investigate the first non-zero eigenvalue of the supra-Laplacian of a node-aligned multiplex network. From the interlacing results of the previous section, we know that 
\begin{equation}
\bar{\mu}_2 \leq {\tilde{\mu}}_{a_2}
\label{mubound1}
\end{equation}
and that
\begin{equation}
\bar{\mu}_2 \leq m.
\label{mubound2}
\end{equation}
$m$ is always an eigenvalue of the supra-Laplacian, so, we can look for the condition under which $\bar{\mu}_2=m$ holds. By combining equations (\ref{mubound1}) and (\ref{mubound2}), we arrive to the conclusion that if $m\geq {\tilde{\mu}}_{a_2}$, then $\bar{\mu}_2\neq m$. On the other hand, we can approximate $\bar{\mu}_2$ as
\begin{equation}
\bar{\mu}_2=\mu_2+\Delta\mu_2,
\end{equation}
where $\mu_2$ is the eigenvalue of $\mathcal{L}$. We have
\begin{equation}
\Delta\mu_2\approx \sum_{i<j}c_{ij}(x_i-x_j)^2,
\label{muapprox}
\end{equation}
where $x_i$ are the elements of the eigenvector $x$ associated to $\mu_2$. Because of the structure of $C$ and $x$, it results
\begin{equation}
\Delta\mu_2\approx m-1
\end{equation}
for a node aligned multiplex. Thus, since $m$ is always an eigenvalue of $\bar{M}$, for that approximation to be correct, the following condition must hold
\begin{equation}
\mu_2 +m -1 <m,
\end{equation}
from which we can conclude that if $\mu_2<1$ then $\bar{\mu}_2\neq m$.\\
In summary, we have that, if $\tilde{\mu}_{a2}<m$ or $\mu_2<1$ then $\bar{\mu}_2\neq m$, but the converse is not true in general.

\section{Applications}

\subsection{Dynamical processes: epidemic spreading}

An important application are the dynamical consequences of the interlacing properties on both adjacency and Laplacian matrices (see Section~\ref{sec:spectral} and Ref.~\cite{ruben}). Here, as an example, we show the SIS epidemic spreading on the top of a multilayer network and the comparison with the aggregate network. Such dynamical process is based on the contact between individuals, or nodes, which can be infected or susceptible to the disease. Infected nodes, also called spreaders, spread the disease to its neighbors inside a time windows with probability $\beta$ and recover from it with probability $\mu$. Considering a discrete time approach, the Markov chain that formalizes this processes can be formally written by the iterative equation
\begin{equation} \label{sis_multilayer}
 p_i(t+1) = \beta \sum_j \bar{\mathcal{A}}_{ij} p_j(t) - \mu p_i(t),
\end{equation}
where $p_i(t)$ is the probability of the node-layer pair $i$ be infected at time $t$, $\bar{\mathcal{A}}_{ij}$ are the elements of the supra-adjacency matrix $\bar{\mathcal{A}}$,  while $\beta$ and $\mu$ are the infection and recovery probabilities, respectively. Such model consider the inter-layer and intra-layer as equal, which is a special case of the model presented in~\cite{multispread}. The critical point can be obtained by the first order approximation of Eq.~(\ref{sis_multilayer}) on its stationary regime, yielding
\begin{equation}
 \beta_c = \frac{\mu}{\lambda_n(\bar{\mathcal{A}})},
\end{equation}
where $\lambda_n(\bar{\mathcal{A}})$ is largest eigenvalue of the supra-adjacency matrix $\bar{\mathcal{A}}$ (see Eq.~(\ref{supradjacency})). From the interlacing properties
\begin{equation}
 \lambda_{n_{\alpha}}(\mathbf{A}^{\alpha}) \leq \lambda_n(\bar{\mathcal{A}}),
\end{equation}
Hence, the critical value $\beta_c$ is bounded by the individual critical values and it is always lower or equal to the lowest individual layer critical value. In addition, observe that when the effective multiplexity, $z \approx 0$ in Eq.~(\ref{eq:z}), the approximated leading eigenvalue of the multilayer supra-adjacency is given by the $\bar{\lambda} = \max \{\lambda(\bar{\mathcal{A}})\}$. Furthermore, exploiting the network of layers spectra, 
\begin{equation}
 \lambda_m \leq \lambda_n (\bar{\mathcal{A}}), 
\end{equation}
where $\lambda_m$ is the largest eigenvalue of the network of layers, whose matrix is given by equation~(\ref{network_of_layers}), implying another constraint to the critical point. In other words, the critical point of the network of layers bound from above the critical point of the multilayer.

Contrasting with the first model, now we consider a spreading process on the aggregate network, Eq.~(\ref{aggregate}), hence
\begin{equation} \label{sis_aggregated}
 p_i(t+1) = \beta \sum_j \tilde{a}_{ij} p_j(t) - \mu p_i(t),
\end{equation}
where $p_i(t)$ is the probability of the node $i$ be infected at time $t$, $\tilde{a}_{ij}$ are the elements of the aggregated adjacency matrix $\tilde{\mathbf{A}}$, $\beta$ is the infection probability and $\mu$ is the recovery probability. Observe that such process is different from the spreading described on Eq.~(\ref{sis_multilayer}), in which each node can infect its neighbors on any layer. On the other hand, in Eq.~\ref{sis_aggregated} each supra-node chooses a layer with uniform probability, than spreads the disease to all neighbors in that layer. Moreover, the critical point can be obtained using the same arguments as before, yielding to
\begin{equation}
 \tilde{\beta_c} = \frac{\mu}{\lambda_n(\tilde{A})},
\end{equation}
where $\lambda_n(\tilde{\mathbf{A}})$ is largest eigenvalue of the aggregated adjacency matrix. Once again, for the interlacing results we have
\begin{equation}
 \tilde{\beta_c} \geq \beta_c.
\end{equation}
Such result imply that the the spreading process on the multilayer structure is more efficient, or in the worst case as efficient as, than the process on the aggregate network~\cite{ruben}.

The results of this section was formerly presented in~\cite{ruben}. In addition, it is noteworthy that a more complete model is proposed in~\cite{multispread}, which consider the activity of the nodes and different spreading probabilities for the intra-layer and inter-layer edges. However, here we show the simplest cases, similar to the ones exposed in~\cite{multispread}, in order to be more didactic. In spite of that, the examples shown here exemplify the importance of considering the multilayer structure and the role of the aggregated network and the network of layers.

\subsection{Real-world multilayer networks}

In order to evaluate real-world multilayer structures we study some networks available at~\emph{http://deim.urv.cat/~manlio.dedomenico/data.php}. We separate them into three different categories: (i) transportation networks; (ii) biological networks and (iii) social networks. We evaluate the maximum of the individual layer eigenvalues and the eigenvalue of the supra-adjacency matrix $\bar{\mathcal{A}}$. Moreover, the approximations of the leading eigenvalues are also computed for comparison. Table~\ref{tab:structure} presents the results. Contrasting with monoplex systems, instead of one type of relationship, here we have $m$ different types and also the connections between different layers. The average of the matrix $\mathcal{A}$ contains information about the relationship inside each layer, whereas the average of $\mathcal{C}$ summarizes the relations between layers, i.e., between a given structure in two different contexts.

Regarding the networks studied here, we observe that biological networks tend to be sparser than social nets, specially considering the inter-layer relations. In addition, observe that there is a relationship between the average of the matrix $\mathcal{C}$ and the effective multiplexity $z$. For most of the networks, the first order approximation is accurate. However, some networks are better approximated by the second order approximation, for instance the CS. Furthermore, among all networks analyzed the only one that presented a poor approximation is the EU air transportation network, which can be explained by the high density of inter-layer couplings compared with the density of intra-layer connections. 

\begin{landscape}
\begin{center}
\begin{table}[t!] 
\centering
% \scalefont{0.8}
\caption{Properties of real multilayer networks.}
\begin{tabular}{|c|c|c|c|c|c|c|c|c|c|c|c|c|}
\hline
& & \multicolumn{7}{c|}{Structure} & \multicolumn{2}{c|}{Approximation} & References\\
 \hline
 & Multilayer & N & m & $\langle \mathcal{A} \rangle$ & $\langle \mathcal{C} \rangle$ & $\langle \bar{\mathcal{A}} \rangle$ & $\max \{ \lambda(\bar{\mathcal{A}}) \}$ & $\max_{\alpha} \{ \lambda(A_{\alpha}) \}$ & $\bar{\lambda} = \lambda + \frac{z}{\lambda}$ & $z$ &  \\
\hline
\multirow{2}{*}{\vspace{-0.75cm}\hspace{0.4cm}\begin{rotate}{90}
\hbox{\scalefont{0.8} Transp.}
\end{rotate}\hspace{0.2cm}}&London transport	&	$369$	&	$3$	&	$2.211$	&	$0.155$	&	$2.366$	&	$3.787$	&	$3.782$	&	$3.786$	&	$0.004$	&  \cite{DeDomenico2014}\\
 & EU air transportation	&	$450$	&	$37$	&	$3.528$	&	$11.417$	&	$14.945$	&	$30.274$	&	$19.315$	&	$19.690$	&	$7.226$	& \cite{Cardillo}\\
\hline
 \multirow{8}{*}{\vspace{-1.5cm}\hspace{0.4cm}\begin{rotate}{90}
\hbox{Biological}
\end{rotate}\hspace{0.2cm}} & C. elegans connectome	&	$279$	&	$3$	&	$7.855$	&	$1.894$	&	$9.748$	&	$21.187$	&	$21.005$	&	$21.099$	&	$1.959$	& \cite{Chen2006}, \cite{DeDomenico2014b}	\\
 & Danio Rerio genetic	&	$155$	&	$5$	&	$1.900$	&	$0.367$	&	$2.267$	&	$4.941$	&	$4.391$	&	$4.484$	&	$0.408$	& \cite{Stark2006}, \cite{DeDomenico2014b}\\
 & Hepatitus C genetic	&	$105$	&	$3$	&	$1.938$	&	$0.419$	&	$2.357$	&	$9.180$	&	$8.888$	&	$9.016$	&	$1.139$	& \cite{Stark2006}, \cite{DeDomenico2014b}\\
 & Homo genetic	&	$18222$	&	$7$	&	$8.570$	&	$1.516$	&	$10.086$	&	$119.311$	&	$119.286$	&	$119.308$	&	$2.541$	& \cite{Stark2006}, \cite{DeDomenico2014b}\\
 & Human Herpes4 genetic	&	$216$	&	$4$	&	$1.847$	&	$0.444$	&	$2.291$	&	$12.484$	&	$12.329$	&	$12.453$	&	$1.530$& \cite{Stark2006}, \cite{DeDomenico2014b}\\
 & Human HIV1 genetic	&	$1005$	&	$5$	&	$2.115$	&	$0.393$	&	$2.508$	&	$16.427$	&	$16.227$	&	$16.364$	&	$2.226$	& \cite{Stark2006}, \cite{DeDomenico2014b}\\
 & Oryctolagus genetic	&	$144$	&	$3$	&	$1.808$	&	$0.093$	&	$1.901$	&	$8.832$	&	$8.832$	&	$8.832$	&	$0.000$	& \cite{Stark2006}, \cite{DeDomenico2014b}\\
 & Xenopus genetic	&	$461$	&	$5$	&	$1.935$	&	$0.488$	&	$2.423$	&	$6.203$	&	$6.093$	&	$6.189$	&	$0.583$	& \cite{Stark2006}, \cite{DeDomenico2014b}\\
\hline
 \multirow{8}{*}{\vspace{0cm}\hspace{0.4cm}\begin{rotate}{90}
\hbox{Social}
\end{rotate}\hspace{0.2cm}}  & CKM Physicians Innovation	&	$246$	&	$3$	&	$4.065$	&	$1.884$	&	$5.950$	&	$8.125$	&	$6.703$	&	$6.998$	&	$1.975$	& \cite{Coleman1957}\\
 & CS Aarhus	&	$61$	&	$5$	&	$5.536$	&	$2.929$	&	$8.464$	&	$11.522$	&	$10.220$	&	$10.476$	&	$2.612$	& \cite{Magnani2013} \\
 & Kapferer Tailor Shop	&	$39$	&	$4$	&	$7.360$	&	$2.893$	&	$10.253$	&	$14.704$	&	$14.052$	&	$14.261$	&	$2.930$	& \cite{Kapferer1972}\\
 & Krackhardt High Tech	&	$21$	&	$3$	&	$7.746$	&	$2.000$	&	$9.746$	&	$14.801$	&	$14.542$	&	$14.680$	&	$2.000$	& \cite{Krackhardt1987}\\
 & Lazega Law Firm	&	$71$	&	$3$	&	$15.670$	&	$1.991$	&	$17.660$	&	$24.194$	&	$23.868$	&	$23.952$	&	$1.994$	& \cite{Snijders2006} \\
 & Vickers Chan 7th Graders	&	$29$	&	$3$	&	$11.908$	&	$2.000$	&	$13.908$	&	$18.426$	&	$18.070$	&	$18.181$	&	$2.000$	& \cite{Vickers1981} \\
\hline
\end{tabular}
\label{tab:structure}
\end{table}
\end{center}
\end{landscape}

\section{Conclusion}

The last years of research have just started to show that interconnected networks exhibit specific structural and dynamical properties that cannot be directly deduced from isolated networks. In order to gain understanding of such a system, a complete new toolbox is needed. On the other hand, such a new framework cannot be a naive extension of what has been developed for isolated, single layered, networks: we need that those tools be adapted to particular questions posed by interconnected networks. It is our conviction that the best way to tackle the problems ahead is to came back to the very basic concepts of graph theory and to build on them. The supra-adjacency matrix and the supra-Laplacian are examples of such basic objects, and the specific structural features of the interconnected system are reflected in them. In this way, the rigorous study of these objects, as well as of their spectral properties, is likely to lead us to the correct understanding of the systems under study. Additionally we presented two applications, firstly the difference an epidemic spreading process that takes place on top of a multilayer or the aggregated network. Secondly, we have shown that perturbation theory is accurate enough when it comes to approximate the eigenvalue of a multilayer structure using the dominant (or co-dominant) layers.

\section{Acknowledgements} 

EC was supported by the FPI program of the Government of Aragon, Spain. This work has been partially supported by MINECO through Grant FIS2011-25167 to YM, Comunidad de Arag\'on (Spain) through FENOL to EC and YM; and by the EC FET-Proactive Project PLEXMATH (grant 317614) to YM. FAR and GAF thank Fapesp and CNPq for financial support give to this research.

\bibliography{biblioMQ}

\end{document}